\documentclass[pdflatex,sn-aps]{sn-jnl}

\usepackage{graphicx}%
\usepackage{multirow}%
\usepackage{amsmath,amssymb,amsfonts}%
\usepackage{amsthm}%
\usepackage{mathrsfs}%
\usepackage[title]{appendix}%
\usepackage{xcolor}%
\usepackage{textcomp}%
\usepackage{manyfoot}%
\usepackage{booktabs}%
\usepackage{algorithm}%
\usepackage{algorithmicx}%
\usepackage{algpseudocode}%
\usepackage{listings}%

\usepackage{import}
\usepackage{tikz}
\usetikzlibrary{math}
\usetikzlibrary{calc}
\usetikzlibrary{decorations.pathreplacing}

\usepackage[normalem]{ulem}

\usepackage[caption=false]{subfig}

\newcommand{\be}{\begin{equation}}
\newcommand{\ee}{\end{equation}}
\newcommand{\ba}{\begin{eqnarray}}
\newcommand{\ea}{\end{eqnarray}}

\begin{document}

\title{What is the Quark-Gluon Plasma made of?}

\author{\fnm{Berndt} \sur{M\"uller}}\email{bmueller@duke.edu}

\affil{\orgdiv{Department of Physics}, \orgname{Duke University}, \city{Durham}, \postcode{27708}, \state{NC}, \country{USA}}

\abstract{This article surveys our present understanding of the internal structure of the fully developed quark-gluon plasma at temperatures outside the crossover region. The theoretical part of the review covers perturbative and nonperturbative approaches to quark-gluon plasma structure, in particular, hard-thermal loop effective theory, lattice QCD and the functional renormalization group. The phenomenological part of the review scrutinizes the information that has been derived from bulk observables and hard probes in relativistic heavy ion collisions in terms of how it informs our knowledge about the structure of the quark-gluon plasma. The final section lists possible avenues for future progress.}




\maketitle

\section{Introduction}\label{sec:Intro}

Historically, two separate but converging paths of reasoning led to the conjecture that strongly interacting matter at the highest energy densities is a quark-gluon plasma (QGP). One, initiated by the seminal contribution of Collins and Perry \cite{Collins:1974ky} was based on the concept of asymptotic freedom, i.~e.\ that the strong interactions become weak at short distance, and therefore quarks and gluons become unbound when the (energy or particle) density becomes sufficiently high. The other, is based on the insight of Cabbibo and Parisi \cite{Cabibbo:1975ig} who postulated that Hagedorn's ``limiting'' temperature is the signal of a phase transition to a new state of matter with unspecified properties. 

Fifty years of experimental and theoretical investigation have shown that both conjectures are only partially right, in a complementary way. The high-temperature or high baryon density phase of hadronic (strongly interacting) matter, indeed, has to be understood as matter described by interacting quarks and gluons, but the interaction is so strong that the constituents of this new state of matter are not appropriately described as ``nearly free'' quarks and gluons. The term ``strongly coupled quark-gluon plasma'' is now customarily used to designate the nature of this new phase of matter \cite{Gyulassy:2004zy}. This review outlines the sometimes convoluted path guided by experimental and theoretical insights that led to our current understanding of ``what the QGP is made of'' and suggests some directions for future exploration. In doing so, I will mostly focus on the ``fully developed'' QGP beyond the transition region near the phase boundary defined by the gradual disappearance of the chiral condensate, i.~e.\ on the temperature range $T > T_c+3\Delta T \approx 200$ MeV, where $\Delta T \approx 15$ MeV is the width of the peak in the chiral susceptibility \cite{Borsanyi:2020fev}, except in Subsection \ref{sec:fRG}, which is devoted to the physics of this transition region.

Section \ref{sec:HTL} describes how our evolving insights into thermal perturbation theory have shaped our understanding of the quark-gluon structure of the QGP. Section \ref{sec:NPMethods} covers some nonperturbative approaches that have contributed to our understanding of the structure of the QGP: Subsection \ref{sec:LGT} reviews results from lattice gauge theory; Subsection \ref{sec:fRG} summarizes insights from functional renormalization group techniques (fRG) with respect to the structure of the QGP in the transition between QGP and hadronic matter. Section \ref{sec:Probes} contains a brief discussion of the connection between observables and the structural properties of the QGP. Sections \ref{sec:Soft} and \ref{sec:Hard} address the state of our knowledge of QGP structure derived from experimental data involving hard and soft probes. The final Section \ref{sec:Outlook} contains a summary and an outlook on open questions for future experimental and theoretical investigation.

There are many aspects of the physics of hot and dense QCD matter that are not covered this review. One of them is the structure of the QGP at high net-baryon density, especially the question whether a critical point exists in the QCD phase diagram. The best current estimates for the location of a critical point lie in the vicinity of $(\mu_B,T) \approx (600~{\rm MeV}, 100~{\rm MeV})$ \cite{Hippert:2023bel,Clarke:2024ugt,Borsanyi:2025dyp}. Another concerns possible new phases of quark matter in the high net-baryon density regime, such as quarkyonic matter \cite{McLerran:2008ua} or an inhomogeneous moat regime \cite{Fu:2024rto}. The review also does not attempt to survey models of the QGP-hadron transition.

The article is intended as a snapshot of our current understanding of ``what the QGP is made of,'' not as a faithful historical account of the developments that led to this understanding. Given the vast literature relevant to the topic of this article, the selection of references by necessity had to be incomplete and sometimes biased. In most cases the interested reader may best start from the most recent citation or from cited review articles to get a more complete view of the published literature and the historical development of the relevant ideas and concepts.

\section{Thermal Perturbation Theory}\label{sec:HTL}

As mentioned above, originally the pervasive vision for the QGP, based on the concept of asymptotic freedom, was that of a weakly coupled ``gaseous'' plasma. The prevalent quasiparticle modes in such a weakly coupled QGP are plasma-modified quarks and gluons. Gluons in a thermal plasma populate two transverse modes and a collective longitudinal mode, the plasmon \cite{Klimov:1982bv,Weldon:1982aq,Pisarski:1989cs}; quarks and antiquarks each have two elementary helicity modes and two reverse-helicity``plasmino'' modes, describing quark holes in the QGP {\cite{Weldon:1989ys,Pisarski:1989wb}. The plasmon and plasmino modes are exponentially suppressed at large momentum. 

It was recognized early on \cite{Kapusta:1979fh,Linde:1980ts} that perturbation theory in a thermal QGP is plagued with infrared divergences. Most of these can be avoided by the introduction of static and dynamical screening of gluon fields, with the exception of the screening of static chromomagnetic modes, which is genuinely nonperturbative \cite{Philipsen:1994ic}. Hard-thermal loop (HTL) perturbation theory \cite{Braaten:1989mz,Taylor:1990ia,Blaizot:2001nr} provides for a formal framework, built on dynamically screened quark and gluon modes, that makes it possible to calculate many near-equilibrium properties of the QGP order by order in the gauge coupling constant $g(T)$. Some quantities that are sensitive to static chromomagnetic fields, such as the color conductivity of the QGP \cite{Selikhov:1993ns}, depend logarithmically on the nonperturbative chromomagnetic mass.

Over the past two decades, an increasing number of thermal QGP properties has been calculated at next-to-leading order (NLO) in $g(T)$. Examples of such properties are the chromoelectric \cite{Rebhan:1993az,Arnold:1995bh,Laine:2019uua,Ekstedt:2023oqb} and chromomagnetic \cite{Bieletzki:2012rd} screening mass, the color conductivity \cite{Arnold:1999uy}, the broadening kernel $C(q)$ \cite{Caron-Huot:2008zna}, the transverse momentum diffusion constant $\hat{q}$ for energetic partons often called the jet quenching parameter \cite{Ghiglieri:2015ala}, and the QGP shear viscosity $\eta$ \cite{Ghiglieri:2018dib,Danhoni:2024ewq}. Even a fully nonperturbative calculation of the collision kernel  calculation has been published \cite{Schlichting:2021idr,Moore:2021jwe}.

The NLO corrections for the transport coefficients $\hat{q}$ and $\eta$ are found to be large at realistic values of $g(T)$, which questions the usefulness of the HTL expansion. However, the reason for the large difference between the LO and NLO results for $\hat{q}$ and $\eta$ can be understood \cite{Muller:2021wri} as a consequence of the radiative (NLO) corrections to the collision kernel $C(q)$. The leading-order HTL result $C(q) \propto q^{-2}$ for small $q$ itself corresponds to a very large, in fact infinite, reduction of the divergent unscreened perturbative result $C(q) \propto q^{-4}$. This unphysical suppression of small-angle scattering is corrected at NLO by the inclusion of radiative processes leading to the dependence $C(q) \propto q^{-3}$ at small momentum transfers \cite{Caron-Huot:2008zna}. The fully nonperturbative result \cite{Schlichting:2021idr} does not qualitatively change this behavior. This raises the hope that corrections beyond NLO for dynamical quantities of physical interest may be modest.\footnote{The detailed structure of color screening -- together with chiral symmetry restoration a defining feature of the QGP -- has been found to be quite complex \cite{Bazavov:2020teh}. We will discuss this aspect of the QGP in more detail in Subsection \ref{sec:LGT}. Here we only note that the Debye screening predicted by NLO perturbation theory provides a semi-quanitative description in the phenomenologically most relevant distance region $0.3~{\rm fm} \leq 0.6~{\rm fm}$ for $T > 300$ MeV \cite{Bazavov:2018wmo}.}

At strong coupling neither gluons nor quarks form well-defined quasiparticles. This can be shown explicitly in the exactly solvable model of ${\cal N}=4$ super-Yang-Mills (SYM) theory at strong 't Hooft coupling, where cut contributions from colored quasiparticles are absent from the thermal spectral functions \cite{Teaney:2006nc}, but the phonon mode is found to be well defined and only weakly damped at momenta $|{\bf k}| < 3T$ \cite{Baier:2007ix}. (For an analysis of the spectral functions in the SYM theory at intermediate 't Hooft coupling see \cite{Casalderrey-Solana:2018rle}.)

\begin{figure}[htb]
\centering
	\includegraphics[width=0.45\linewidth]{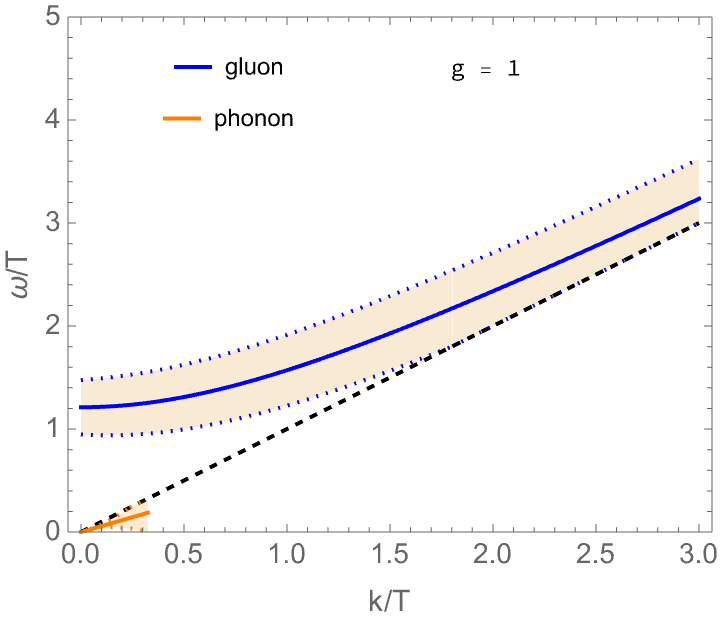}
	\hspace{0.03\linewidth}
	\includegraphics[width=0.45\linewidth]{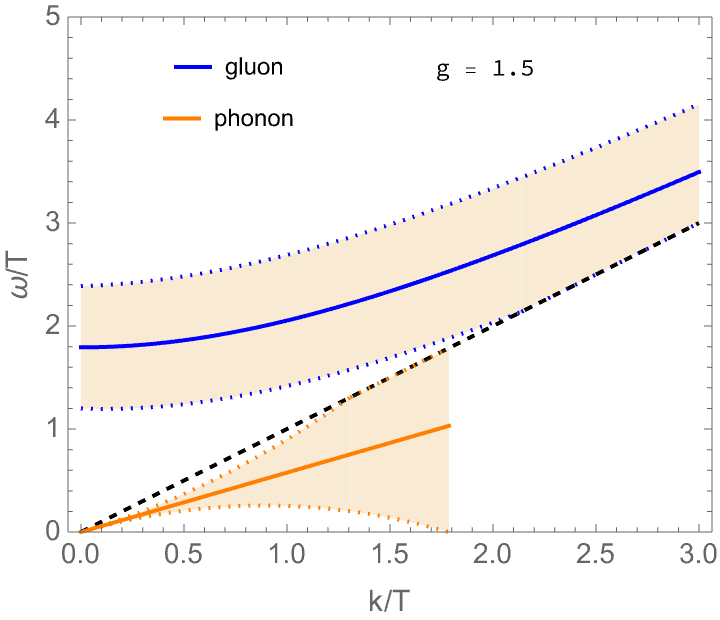}
	\hspace{0.03\linewidth}
	\includegraphics[width=0.45\linewidth]{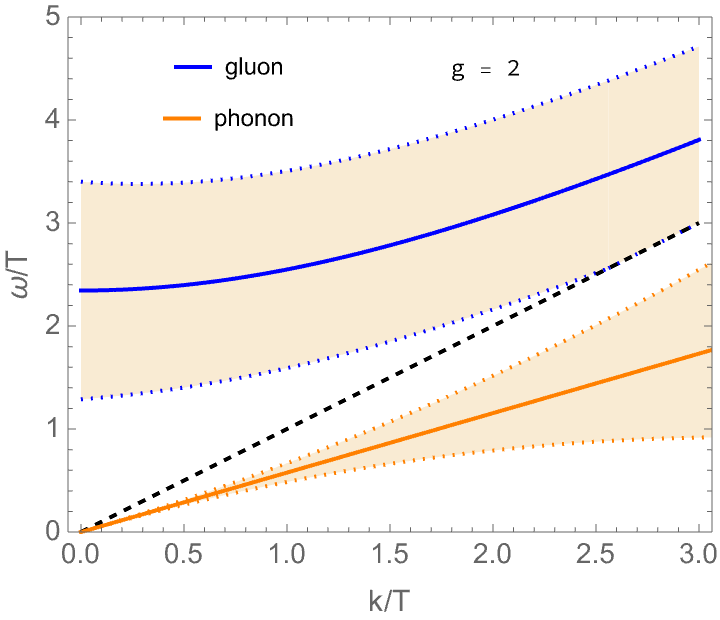}
	\hspace{0.03\linewidth}
	\includegraphics[width=0.45\linewidth]{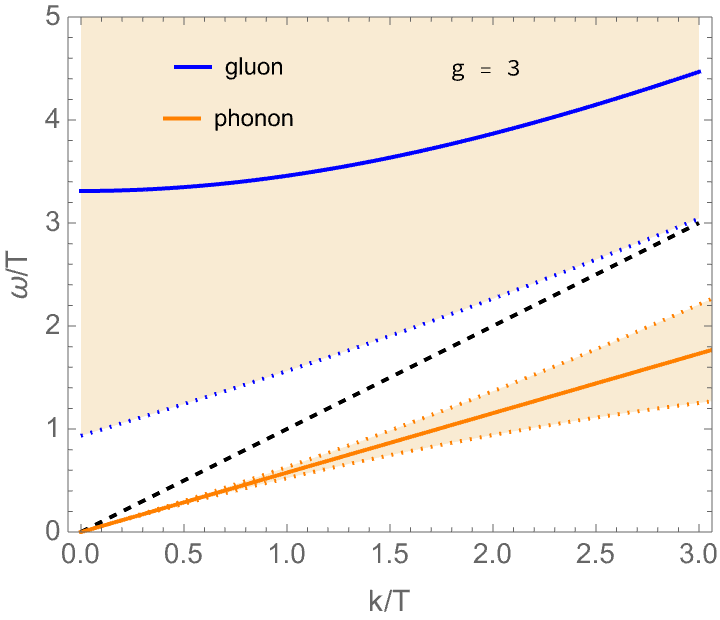}
\caption{Quasiparticle structure of the quark-gluon plasma in NLO perturbation theory at gauge coupling $g=1$ (top left), $g=1.5$ (top right), $g=2$ (bottom left) and $g=3$ (bottom right). The time-like quasiparticle above the dashed $\omega=k$ line is the in-medium gluon (plasmon); the space-like quasiparticle below the $\omega=k$ line is the phonon. The shaded areas indicate the quasiparticle widths. The phonon branch is only shown for wave numbers below the inverse sound attenuation length in the QGP. Clearly, the weakly coupled QGP is dominated by in-medium gluons; while the phonon emerges as a well-defined collective mode in the thermal momentum range in the strongly coupled QGP heralding its characteristic property of nearly ``perfect'' fluidity.}
\label{fig:Quasiparticles}
\end{figure}

To illustrate the quasiparticle structure of the QGP at different values of the QCD coupling $g$ the gluon and phonon dispersion relations calculated in thermal perturbation theory are shown in Fig.~\ref{fig:Quasiparticles} for $g = 1,\dots,3$. The dispersion relations for gluons and phonons, respectively, have the form
\begin{eqnarray}
{\rm gluon:}\qquad & \omega_{\rm g}(k) \approx \sqrt{\omega_{\rm p}^2+k^2},\qquad
& \gamma_{\rm g}(k) = \frac{g^2N_c}{4\pi} r(g,k) T \\
{\rm phonon:}\qquad & \omega_{\rm ph}(k) = c_s k, \qquad
& \gamma_{\rm ph}(k) = \frac{2\eta}{3sT} k^2,
\end{eqnarray}
where $\omega_{\rm p}$ is the plasma frequency,  $c_s$ is the speed of sound, $\eta/s$ is the kinematic shear viscosity, and $r(g,k)$ is a function of order unity that interpolates between the static limit \cite{Braaten:1990it} and the high-momentum limit \cite{Arnold:1999uy,Ekstedt:2023oqb} of the gluon damping rate. The real part of the quasiparticle energy as function of momentum ${\bf k}$ is shown by solid lines (blue for gluons, red for phonons); the width $\gamma({\bf k})$ is indicated by the orange shaded regions. The phonon mode is only shown for momenta $|{\bf k}| < \ell_{\rm s}^{-1}$, where $\ell_{\rm s}(k) = c_s/\gamma_{\rm k}$ is the sound attenuation length.\footnote{The authors of \cite{Kovtun:2011np} argue that one should also impose the slightly more restrictive constraint $\omega({\bf k}) < \tau_\pi^{-1}$, where $\tau_\pi$ is the shear stress relaxation time.} Thermal gluons are seen to form relatively long-lived quasiparticles at weak coupling ($g=1$) but become increasingly ill-defined modes at stronger coupling. Conversely, at weak coupling the phonon mode is quasiparticle-like only at very small momenta but emerges at increasingly stronger coupling as a well-defined quasiparticle. 

As the figure for $g=3$ (where thermal perturbation theory is not expected to be reliable) shows, the plasma-modified gluon dissolves at very strong coupling into a quasi-continuum, and the phonon mode becomes the most prominent quasiparticle mode. As mentioned above, this transition from broad quasiparticle modes at intermediate coupling to a gauge plasma at strong coupling, in which the phonon is the only well-defined propagating mode, can be rigorously demonstrated in the exactly solvable model of ${\cal N}=4$ SYM theory \cite{Casalderrey-Solana:2018rle}.   

\begin{figure}[hbt]
\centering
	\includegraphics[width=0.45\linewidth]{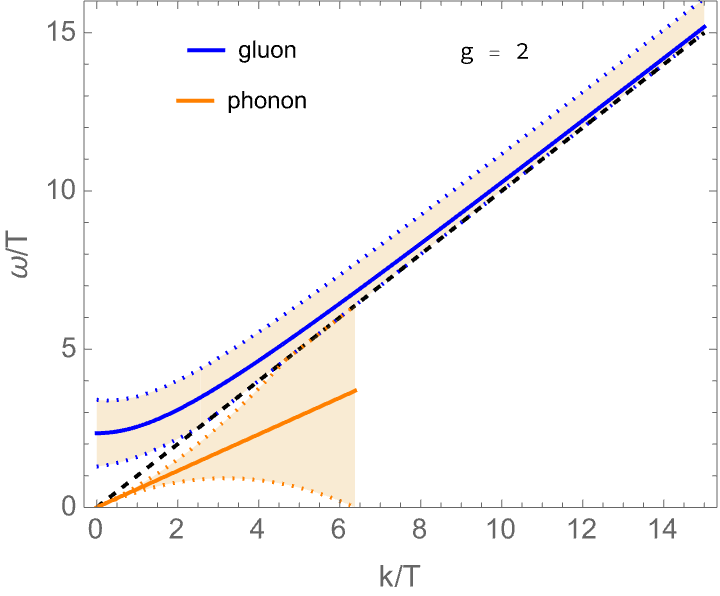}
	\hspace{0.03\linewidth}
	\includegraphics[width=0.45\linewidth]{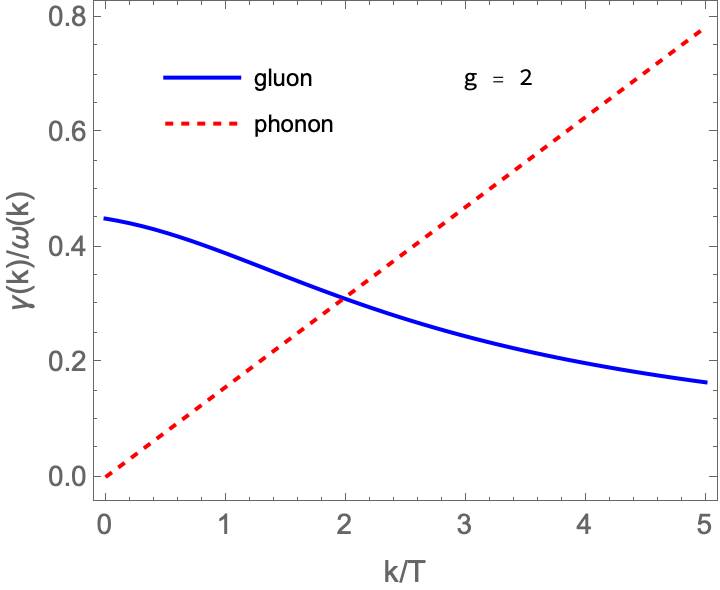}
\caption{{\it Left panel:} Quasiparticle structure of the quark-gluon plasma in NLO perturbation theory at gauge coupling $g=2$ over an expanded momentum range. The gluon mode is seen to become relatively better defined at increasing momentum, whereas the phonon mode is strongly damped for $|{\bf k}| > 6T \approx 2~{\rm GeV}$. {\it Right panel:} The inverse quality factors $Q(k)^{-1} = \gamma(k)/\omega(k)$ for the plasma gluon and phonon modes as function of momentum $k$ at $g=2$. A low value of $Q^{-1}$ ($Q \gg 1$) means that a mode is weakly damped; a high value of $Q^{-1}$ ($Q < 1$) indicates an overdamped mode. The phonon mode is better defined for $|{\bf k}| < 2T$, whereas the gluon mode is better defined for larger momenta.}
\label{fig:QP20}
\end{figure}

A realistic value of $g(T)$ for QGPs formed in the heavy ion collisions is\footnote{We use the value $\alpha_s(1.5~{\rm GeV})= 0.326\pm 0.019$ derived from lattice QCD \cite{Bazavov:2012ka}, together with the thermal scale set as $2\pi T = 1.5$ GeV or $T\approx 250$ MeV.} $g(T) \approx 2$ ($\alpha_s\approx 0.3$), where the gluon modes have not completely dissolved yet, but the phonon mode is already rather well defined for thermal momenta $|{\bf k}| \leq 3T$. An expanded view of the gluon and phonon dispersion relations for $g=2$ is shown in the left panel of Fig.~\ref{fig:QP20} up to $|{\bf k}| \leq 15\,T$. The gluon mode is seen to become better defined at increasing momentum, whereas the phonon mode is strongly damped for ${\bf k}| > 6\,T \approx 1.5~{\rm GeV}$. The right panel of Fig.~\ref{fig:QP20} compares the momentum-dependent relative widths $\gamma(k)/\omega(k)$ of the gluon and phonon modes for $g=2$. The phonon mode is seen to be better defined for $|{\bf k}| < 2T$, whereas the gluon mode has a smaller width for larger momenta. We will discuss the possible implications of this finding in Section \ref{sec:LGT}.

This leads to the following picture of the structure of the QGP based on HTL perturbation theory: At momenta larger than approximately 1 GeV/c the QGP can be understood as a medium composed of quark- and gluon-like quasiparticles with a typical lifetime of order $\gamma^{-1} \approx T^{-1} \sim 0.5~{\rm fm/c}$. The interactions among these quasiparticles is strongly enhanced by radiative processes, which result in large inelastic cross sections and a short mean-free path for momentum transport. This, in turn, manifests itself in an anomalously small kinematic shear viscosity $\eta/s \approx 0.1-0.2$, which implies that phonons are the dominant dynamical mode\footnote{This statement only concerns the effectiveness of the medium response. Thermodynamic quantities weight the quasiparticle states with their degeneracy (16 for gluons, 12 for each quark flavor) while phonons constitute a single scalar mode.} at momenta below $2\, T \approx 0.5$ GeV/c leading to the near ``perfect'' fluid behavior of the QGP that will be discussed further in Section \ref{sec:Soft}.

HTL perturbation theory at three loops (NNLO) has been used to compute the equation of state of the QGP at finite temperature and density \cite{Andersen:2010wu,Andersen:2011sf}. This calculation assumed that QGP thermodynamics can be described by screened quark and gluon quasiparticles interacting via screened color fields. At the NNLO level this description includes all the effects discussed in this section except possible dynamical effects of phonons on the equation of state. As shown in \cite{Andersen:2014dua} the results of this calculation agree well with results from lattice gauge theory, up to uncertainties of scale setting, except in the region near the pseudocritical temperature $T_c$ (see Fig.~\ref{fig:EOS}). This finding lends further support to the picture that the QGP can be described as a medium of short-lived quark and gluon quasiparticles that behaves like a low-viscosity liquid at low frequencies and momenta.

\begin{figure}[hbt]
\centering
	\includegraphics[width=0.44\linewidth]{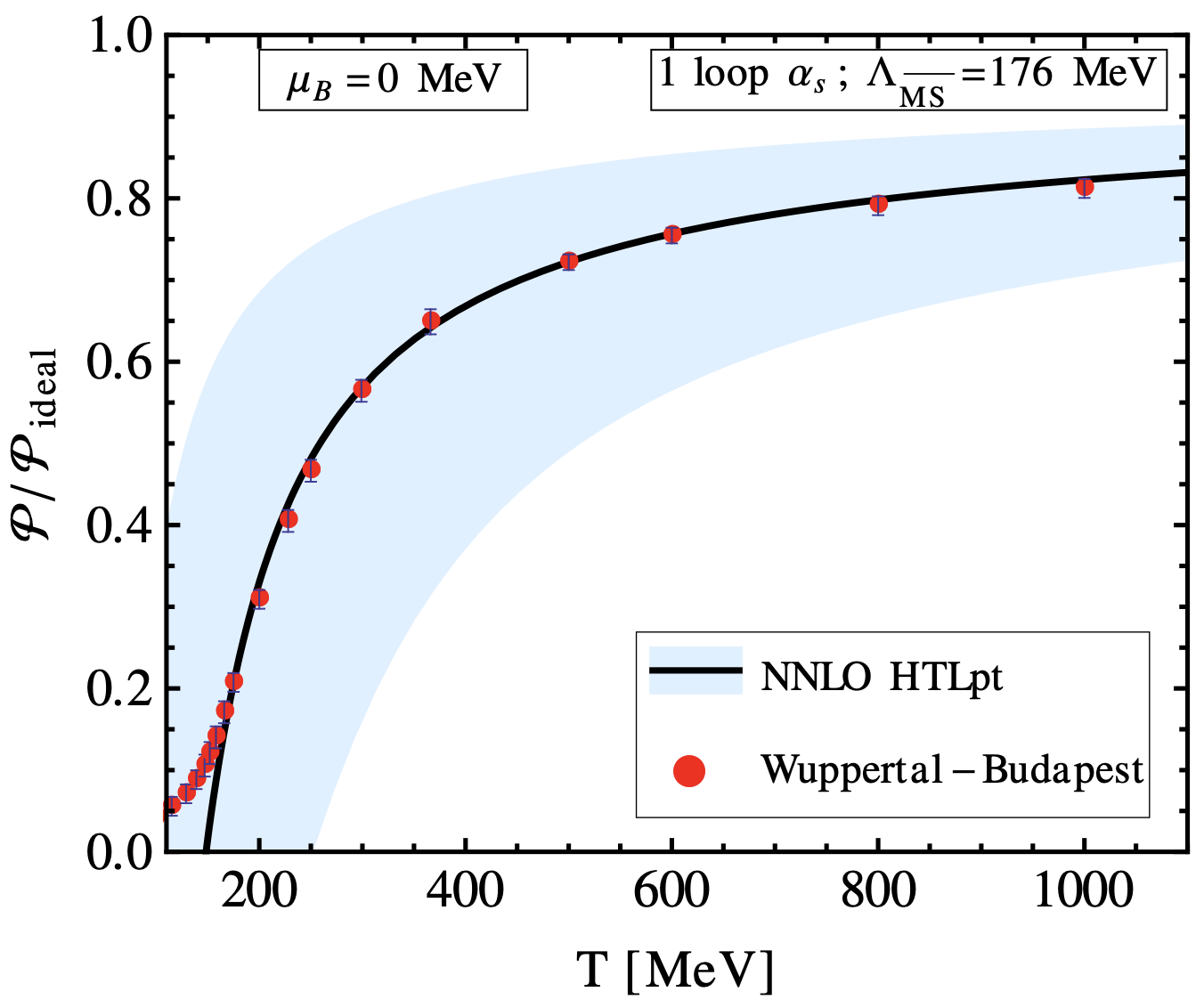}
	\hspace{0.03\linewidth}
	\includegraphics[width=0.45\linewidth]{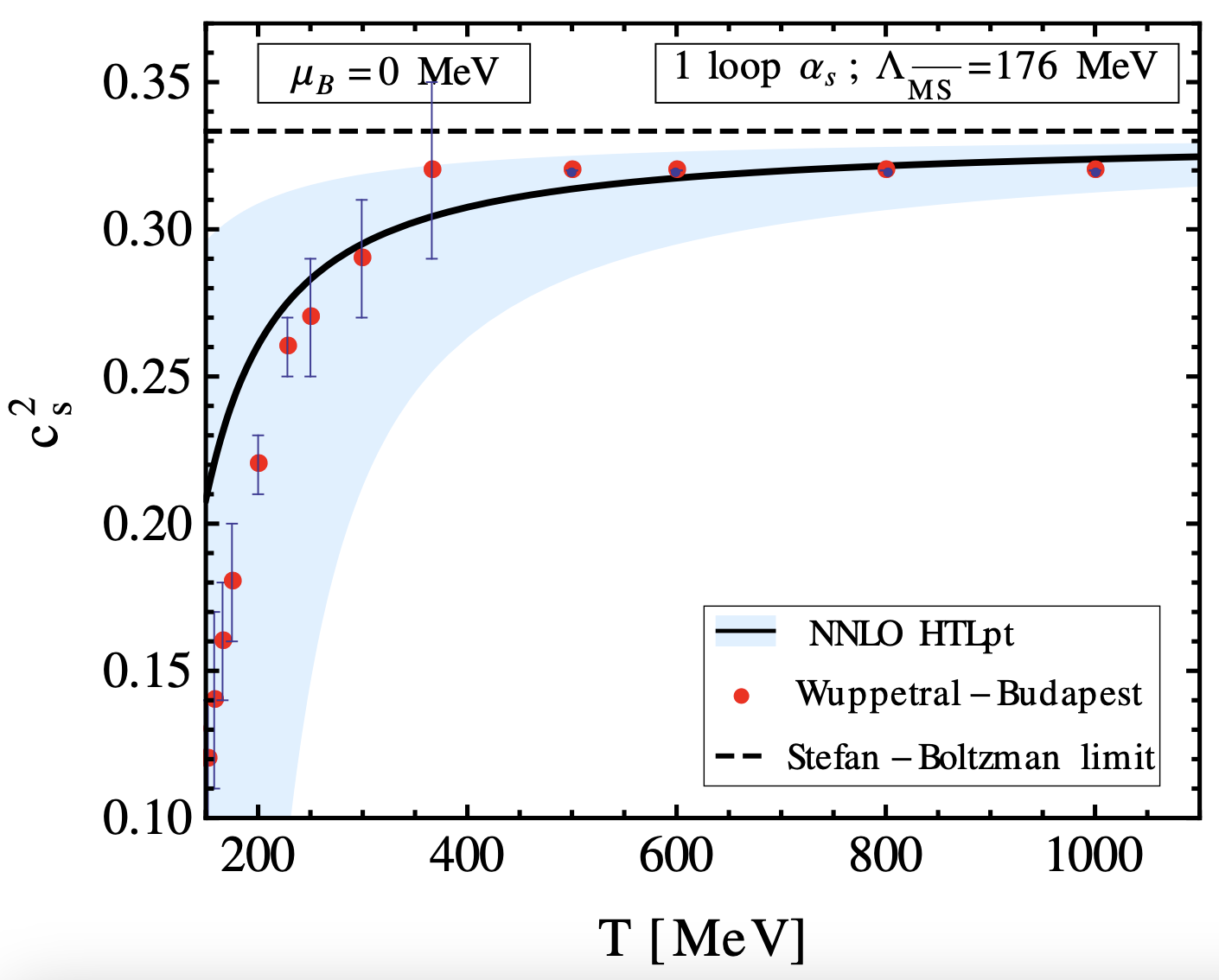}
\caption{{\it Left panel:} Pressure $P$ of the HTL quasiparticle QGP at NNLO normalized to the pressure of a free quark-gluon gas. The solid line shows the pressure for the indicated value of the QCD scale parameter $\Lambda_{\rm QCD}$ at the coupling $\alpha_s(2\pi T)$ in comparison with the lattice QCD results \cite{Borsanyi:2010cj} (red points with error bars). The shaded region indicates the scale setting uncertainty when the scale is varied between $\pi T$ and $4\pi T$. {\it Right panel:} Speed of sound in the HTL quasiparticle QGP at NNLO for the same scale parameter choices as in the left panel. Reproduced from \cite{Andersen:2014dua} with permission by the authors.}
\label{fig:EOS}
\end{figure}

Because HTL perturbation theory is formulated in Minkowski space, it can be used to construct a kinetic theory that describes the evolution of occupation numbers of quark and gluon quasiparticle states \cite{Arnold:2002zm}. This kinetic theory, often referred to as ``AMY'', describes binary scatterings among quasiparticles as well as radiative ($q\to qg, g\to gg$) and absorptive ($qg\to q, gg\to g$) processes involving off-shell quasiparticles.

Another intriguing aspect of the HTL perturbative QGP is the emergence of exponentially growing modes when the thermal distribution is deformed by a momentum anisotropy (see \cite{Mrowczynski:2016etf} for a comprehensive review). These instabilities exist at any gauge coupling and the range of unstable modes grows with increasing anisotropy. It is likely that these modes play a role in the isotropization of the initial gluon distribution \cite{Arnold:2004ti}.

\section{Nonperturbative approaches}\label{sec:NPMethods}

Over the past 50 years many models of the QCD vacuum and hot QCD matter have been proposed and sometimes extensively studied that aim at better understanding of nonperturbative properties of the QGP. Only for one of these methods, lattice gauge theory, has it been demonstrated that rigorous theoretical predictions for many QGP properties with fully controlled errors can be obtained. Insights into ``what the QGP is made of'' from Lattice Gauge Theory will be discussed in Subsection \ref{sec:LGT}. Other approaches either represent models of QCD, such as the instanton liquid model \cite{Schafer:1995pz,Schafer:1996wv}, chromomagnetic monopole liquid  model \cite{Liao:2006ry,Liao:2008pu}, stochastic vacuum model \cite{Antonov:2010tu}, ``semi-QGP'' model of the deconfinement transition \cite{Pisarski:2016ixt}, Dyson-Schwinger equations \cite{Gao:2015kea,Fischer:2018sdj}, and holographic models of QCD \cite{Casalderrey-Solana:2011dxg,Rougemont:2023gfz}. Holographic models have been especially useful as they often yield analytical results for thermodynamic and dynamic properties in the strong coupling limit. We will refer to some of these in Section \ref{sec:Probes} dedicated to experimental probes of the QGP.  An analytical approach that can, in principle, capture all nonperturbative aspects of an interacting quantum field theory and that has been extensively applied to QCD, is the functional renormalization group (fRG) \cite{Wegner:1972ih,Polchinski:1983gv,Wetterich:1992yh,Polonyi:2001se,Gies:2006wv,Dupuis:2020fhh}. The fRG approach is particularly well suited to describe the transition from the elementary degrees of freedom of QCD (quarks and gluons) to the effective degrees of freedom at low-energy (hadrons) and to shed light on the structure of the QGP as the temperature approaches the pseudocritical temperature $T_c$ from above. It is important to keep in mind, however, that while being formally exact, in practice, implementations of the fRG approach have applied physically motivated truncations in the infrared domain where QCD is strongly coupled. Insights derived from fRG treatments will be discussed in Subsection \ref{sec:fRG}.

\subsection{Lattice Gauge Theory}\label{sec:LGT}

Lattice gauge theory, in its Euclidean space version, has emerged as a most powerful {\it ab initio} tool for nonperturbative, definitive calculations of quasi-static properties of the QGP. These include the equation of state \cite{Borsanyi:2013bia,HotQCD:2014kol,Ratti:2022qgf} and its derivative quantities, most notably various susceptibilities and the speed of sound. The speed of sound determines the slope of the dispersion relation for the phonon mode. The susceptibilities associated with conserved quantum numbers are related to the density of quasiparticles carrying these quantum numbers in the QGP. However, the Euclidean lattice simulations cannot directly tell us whether these degrees of freedom are associated with well-defined propagating quasiparticle modes for quarks and gluons. That said, it is worth noting that NNLO HTL calculations of the equation of state, speed of sound and baryon susceptibilities, nicely agree with the lattice results after scale setting except in the temperature range near $T_c$ \cite{Andersen:2014dua}.

One main success of thermal lattice QCD is the  calculation of the equation of state as function of $T$ and $\mu_B$ in the range $\mu_B/T \leq 3.5$ with controlled errors \cite{Borsanyi:2021sxv}. An important result is that the transition between hadron matter and the QGP is a smooth crossover over this entire range and constraints on the existence of a critical point that would mark the beginning of a phase transition line have been derived \cite{Borsanyi:2025dyp}.  A recently reported precision determination of the speed of sound of the QGP from the multiplicity dependence of the mean transverse momentum $\langle p_T \rangle$ \cite{CMS:2024sgx} is in excellent agreement with the lattice QCD results, but the theoretical uncertainties of the extracted value are still under debate \cite{Gardim:2024zvi,Gavassino:2025bts}. Lattice results for various susceptibilities also have been favorably compared with experimental data \cite{Borsanyi:2013hza}. For a comprehensive assessment of the experimental verification of many lattice QCT predictions for the thermal properties of the QGP, see Chapter 7 in \cite{Ratti:2021ubw}.

Other important lattice results concern the restoration of chiral symmetry and color screening in the QGP. The lattice results clearly show the rapid disappearance of the vacuum quark condensate for $u$ and $d$ quarks above $T_c$ accompanied by a rapid rise of the light quark susceptibility and a slightly slower rise of the strange quark susceptibility \cite{Borsanyi:2010bp,Bazavov:2011nk}. These quantities match those obtained in the hadron resonance gas (HRG) model below $T_c$ and show quantitative agreement with NNLO HTL-perturbative calculations for $T \gtrsim 250$ MeV \cite{Mogliacci:2013mca,Andersen:2014dua}.

Color screening in thermal lattice QCD is a complex phenomenon with several different regimes \cite{Bazavov:2020teh}. The overall effect of color screening is measured by the expectation value of the Polyakov loop $\langle L \rangle$, which measures the Gibbs factor $e^{-V_Q/T}$ for the static potential energy $V_Q$ of an isolated heavy quark. $\langle L \rangle$ rapidly rises with temperature across $T_c$, indicating quickly diminishing energy in the static chromoelectric field surrounding the heavy quark \cite{Borsanyi:2010bp,Bazavov:2011nk}. On intermediate length scales of order $(gT)^{-1}$ (0.3 fm $\leq r \leq$ 0.6 fm for $T = 300$ MeV) the static chromoelectric force is Yukawa screened with a Debye masss $m_D$ slightly larger than the NLO HTL prediction \cite{Bazavov:2018wmo} while the static chromomagnetic force is more weakly screened in line with expectations from HTL perturbation theory where static chromomagnetic fields remain unscreened \cite{Maezawa:2010vj}.

At longer distances nonperturbative effects cause static chromomagnetic fields to develop a confining mass gap of order $g^2T$. This effect can be incorporated into the perturbative HTL approach by the addition of a three-dimensional action describing the nonperturbative dynamics of static chromoelectric and chromomagnetic gauge fields, a formalism known as electrostatic QCD or EQCD \cite{Braaten:1995jr}. Color screening in this asymptotic region must be computed on the lattice for any value of the gauge coupling $g$ \cite{Hart:2000ha,Bazavov:2018wmo}. Confinement dictates that the static force between two heavy color charges in this domain is transmitted by color singlet bound states of static gluons. In the temperature range of interest ($T \approx 2T_c$) the lightest bound state is found to be formed by two electrostatic gluons \cite{Hart:2000ha}. This state controls the asymptotic range of the force between two static quarks.\footnote{A detailed description of the complex physics in this asymptotic region can be found in Section 3.2.5 of \cite{Bazavov:2020teh}.} However, it is important to bear in mind that these static properties of the QGP may have limited influence on the dynamic phenomena that are associated with experimental observables.

The dynamical response of the QGP can be deduced from Euclidean correlation functions computed on the lattice (or in some other approach) by analytic continuation to real time. Because lattice simulations only yield values at discrete imaginary times, this process is inherently model dependent. Nevertheless, numerous attempts have been made to obtain estimates for various quantities of interest, such as the electrical conductivity \cite{Aarts:2020dda}, the shear viscosity (for quenched QCD) \cite{Meyer:2007ic,Mages:2015rea,Altenkort:2022yhb}, and diffusion coefficients for electrical charge \cite{Aarts:2014nba} and heavy quarks \cite{Altenkort:2023eav,Altenkort:2023oms}. While these results {\it per se} cannot provide direct information about ``what the QGP is made of'', they help constrain phenomenological models and determine parameters in effective theories of the response of the QGP to perturbations. 

At weak coupling, response functions corresponding to transport processes, such as diffusion or viscous flow, exhibit a transport peak at low frequencies \cite{Hong:2010at}. The height of this peak determines the transport coefficient, its width is related to the characteristic damping rate of quasiparticles. With increasing strength of the coupling this peak broadens and its height drops, until it eventually disappears when there are no well-defined thermal quasiparticles associated with the elementary degrees of freedom (see \cite{Casalderrey-Solana:2018rle} for an analysis in the context of the ${\cal N}=4$ SYM theory). Because the transport peak, if it exists, manifests itself at small frequencies, it is usually difficult, if not impractical, to find it by analytic continuation from Euclidean response functions on the lattice \cite{Petreczky:2005nh,Altenkort:2022yhb,Altenkort:2023oms}.

All results obtained by analytic continuation of lattice QCD simulations indicate that the QGP at conditions attainable in relativistic heavy ion conditions is a strongly coupled medium. The results are generally in rough agreement with the predictions of NLO HTL perturbation theory at realistic coupling $g(T) \approx 1.5-2$, i.~e.\ they are consistent with a quasiparticle structure of the QGP as described in Section \ref{sec:HTL}.

\subsection{Functional Renormalization Group}\label{sec:fRG}

The functional renormalization group (fRG) approach is based on a differential equation that governs the evolution of the effective action $\Gamma_k[\Phi]$, where $\Phi$ denotes the relevant set of fields, as the ultraviolet momentum cut-off $k$ is lowered \cite{Polchinski:1983gv,Wetterich:1992yh}. When composite fields arise as effective degrees of freedom in the infrared domain, they can be incorporated into the action $\Gamma_k[\Phi]$ as emergent dynamical fields. In QCD, these emergent degrees of freedom correspond to color-singlet hadrons, which provide the appropriate description of strong interaction physics at low energies. At finite temperature, the interplay between fundamental fields and emergent fields can describe the gradual transition from the QGP to hadronic matter in the thermal region around $T_c$. 

Investigations in the fRG approach of the QGP and its hadronization have focused on a minimal set of fields: Gluons, ghosts, and quarks as fundamental fields, and scalar and isoscalar meson fields as emergent composites \cite{Braun:2014ata}.\footnote{Phonons arise as collective modes. More comprehensive studies, also including composite isovector vector and axial vector fields corresponding to the $\rho$ and $a_1$, have also been pursued \cite{Rennecke:2015eba,Braun:2019aow}.} The scalar and isoscalar fields are associated with the sigma meson ($\sigma$) and pion ($\pi$), respectively; the quark condensate corresponds to the vacuum expectation value $\sigma_0$. The functional equation for $\Gamma_k$ contains terms describing the in-medium gluon, ghost, quark and meson propagators, scale-dependent vertices and an effective potential for the Polyakov loop. Limited information from experiment and lattice gauge theory is used to fine-tune the QCD effective action. 

Gluons and quarks are found develop mass gaps in the infrared, the latter being generated by the quark condensate that breaks chiral symmetry. The quark-gluon coupling gets increasingly screened below a scale $k \approx 700$ MeV, corresponding to a constituent light quark mass $m_{u,d}\approx 350$ MeV, while the quark-meson coupling saturates at a finite value $h\approx 13$ in the infrared \cite{Braun:2014ata}. The structure of the functional evolution equation ensures that there is no double counting of degrees of freedom.

The dependence of dynamical chiral symmetry breaking in QCD on temperature and baryon chemical potential has been studied for this minimal version of the effective action in considerable detail \cite{Fu:2019hdw,Fu:2022gou}. The quark condensate and the chiral quark mass are found to begin melting around $T \approx 100$ MeV and effectively disappear above $T \approx 200$ MeV. This behavior encodes the restoration of chiral symmetry and is in quantitative agreement with the lattice QCD results.\footnote{Note that the chiral quark mass does not capture the infrared limit of the quark self-energy, which gives the non-chiral thermal quark mass in HTL perturbation theory and grows with $T$.}  In contrast, the effective infrared gluon mass begins to grow with $T$ and reaches approximately 750 MeV at $T = 300$ MeV, which is comparable to the thermal gluon mass in NLO HTL perturbation theory \cite{Cyrol:2017qkl}. Meanwhile the effective scalar and pseudoscalar meson masses become degenerate around $T \approx 170$ MeV and then rise rapidly with $T$ implying a rapid decoupling of these fields from the effective action. The meson degrees of freedom begin to play an increasingly large role on thermal scales as the temperature approaches $T_c$ from above and dominate in the infrared domain below $T_c$.

The fRG formalism facilitates the calculation of real-time correlation functions and spectral functions by analytic continuation \cite{Pawlowski:2015mia}. Spectral functions of gluons have been calculated for the pure SU(3) Yang-Mills theory and for full QCD including dynamical quarks at finite temperature  \cite{Haas:2013hpa,Christiansen:2014ypa} and in the vacuum \cite{Cyrol:2018xeq}. A contour plot of the spectral function of transverse gluons at $2.77\, T_c$, see left panel of Fig.~\ref{fig:fRG}, reveals a similar structure as the panel for $g=2$ in Fig.~\ref{fig:Quasiparticles}.  Calculations of the kinematic shear viscosity $\eta/s$ as function of temperature \cite{Haas:2013hpa,Christiansen:2014ypa} are consistent with the results from HTL perturbation theory and lattice QCD, see right panel of Fig.~\ref{fig:fRG} for the pure gauge theory. 

\begin{figure}[hbt]
\centering
	\includegraphics[width=0.44\linewidth]{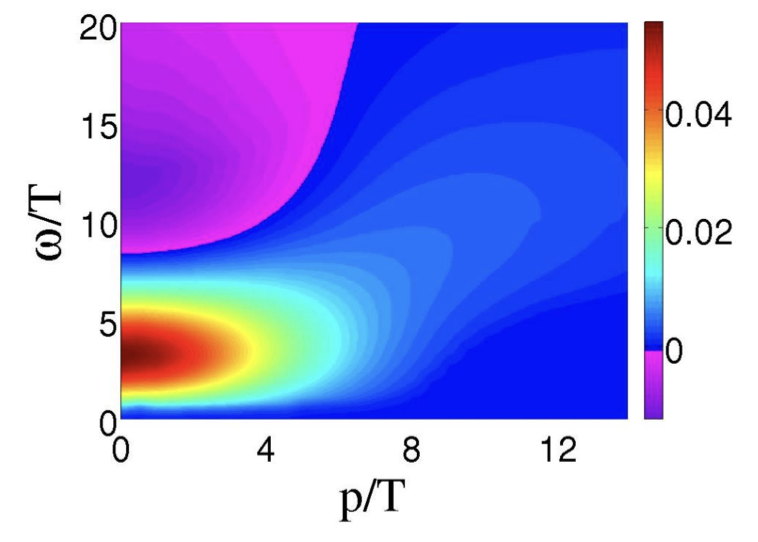}
	\hspace{0.03\linewidth}
	\includegraphics[width=0.47\linewidth]{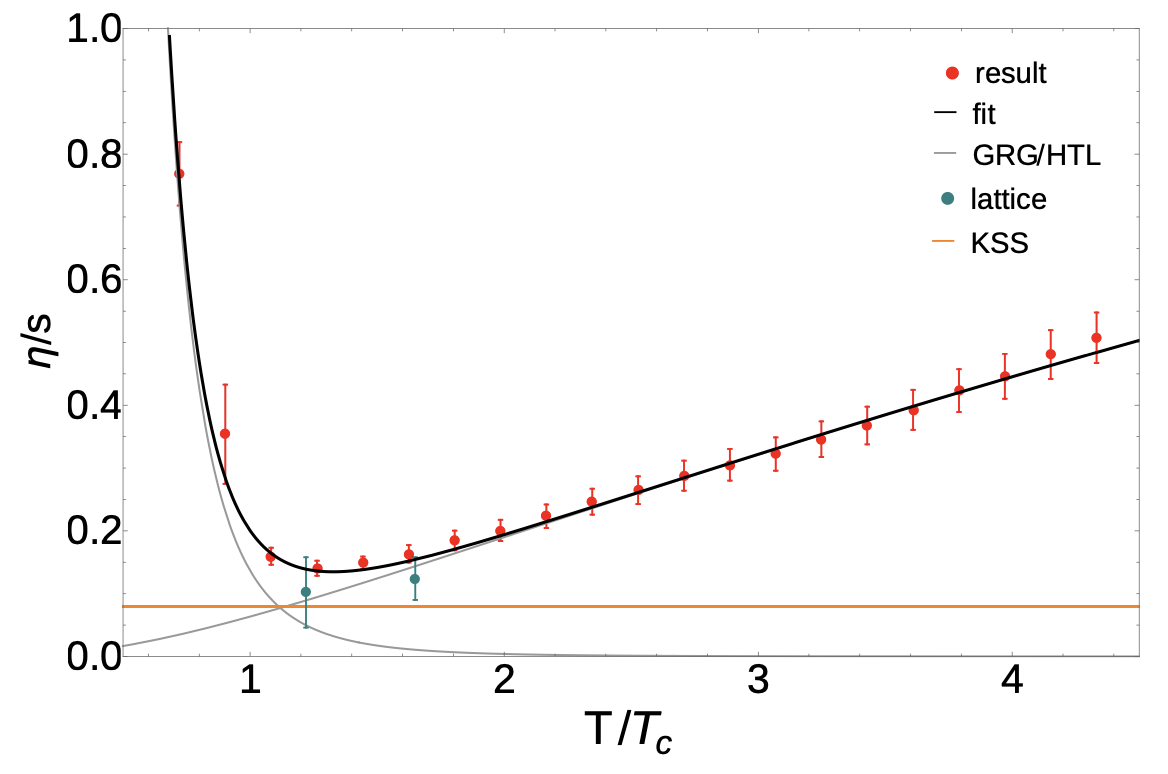}
\caption{{\it Left panel:} Contour plot of the transverse gluon spectral function in quenched QCD in the fRG formalism at $T = 2.77\, T_c$. The in-medium gluon mode appears as a broad resonance starting at the plasmon mass $\omega_p/T \approx 3$ at $p=0$ and approaches the light-cone for momenta $p/T\gtrsim 5$. Reproduced from \cite{Haas:2013hpa} with permission by the American Physical Society. {\it Right panel:} The red dots with error bars show the kinematic shear viscosity $\eta/s$ of pure SU(3) gauge theory calculated in the fRG formalism. The thin lines represent analytical fits inspired by HTL perturbation theory and the glueball resonance gas, respectively. The heavy black solid line is obtained as combination of the two fits. The blue dots with error bars show early lattice results \cite{Meyer:2007ic}. Reproduced from \cite{Christiansen:2014ypa} with permission by the American Physical Society.}
\label{fig:fRG}
\end{figure}

It is worth mentioning in this context that spectral functions of gluon and quark modes in an overoccupied glasma can be computed by means of classical statistical lattice simulations. These exhibit well-defined quasiparticle-like behavior in ($3+1$) dimensions \cite{Boguslavski:2018beu,Boguslavski:2021kdd}, but not in ($2+1$) dimensions \cite{Boguslavski:2021buh}, where all excitations are found to be broad.

\section{Probing the Quark-Gluon Plasma}\label{sec:Probes}

In order to explore the internal structure of the QGP one needs to probe it. Since the QGP is much too short-lived under laboratory conditions to allow for interrogation with external probes, in practice, all probes must be internally generated, either by processes occurring in the initial moment of the collision (``hard'' probes \cite{Satz:1995sa,Satz:1995cg}) or by properties of the QGP itself that manifest themselves in the pattern of particle emission from the bulk matter (``soft'' observables). In the former case, information about the structure of the QGP is carried by the modifications of the hard probes caused by interactions with the QGP, commonly called ``nuclear'' modifications, in the latter, the observed patterns must be traceable to specific transport properties of the QGP. Some physical phenomena, such as jets, can simultaneously act as hard probes and sources of soft observables (e.~g.\ via a jet-induced wake). In both cases, the observables used to analyze the internal structure of the QGP can be related to the response of the QGP to perturbations away from equilibrium.

In many-body theory, the response of the medium to a (small) perturbation is described by linear response theory \cite{Fetter:2012quantum,Hofmann:1976apo}. The response is characterized by a frequency- and momentum-dependent {\it response function} $S(\omega,{\bf k})$ that depends on the quantum numbers  governing the perturbation and encodes the internal structure of the medium. When the medium is weakly coupled or dilute, the response functions are usually dominated by quasiparticles, i.~e.\ excitations characterized by a particle-like dispersion relation $\omega({\bf k})$ with an imaginary part (width) $\gamma({\bf k}) \ll \omega({\bf k})$.  An early systematic survey of experimental QGP signatures in terms of response functions can be found in \cite{Muller:1994rb}.

When the medium is strongly coupled and dense, the response functions generally cannot be interpreted in terms of quasiparticles. However, in certain channels corresponding to the transport of conserved quantities, such as energy and momentum, {\it collective} quasiparticles exist even at strong coupling. A prime example for this phenomenon is provided by the quanta associated with the sound mode, i.~e.\ phonons, which are good quasiparticles only at parametrically long wavelengths in weakly coupled media, but can become the most well-defined quasiparticles on all distances in strongly coupled media as discussed in Section \ref{sec:HTL}.

In the next two sections we will discuss specific experimentally accessible probes and review what the data have so far told us about the internal structure of the QGP. We begin with soft observables in Section \ref{sec:Soft} and then turn to hard probes in Section \ref{sec:Hard}.

\section{Soft Observables of the QGP}\label{sec:Soft}

A connection between the internal structure of the QGP and the bulk observables requires an intimate understanding of the dynamical evolution of the QGP in a heavy ion collision based on transport theory. Most practical formulations of transport theory employ severe simplifications by restricting the description to those aspects of the quantum system that evolve most slowly. The current paradigm distinguishes four sequential stages: (1) Nonlinear dynamics of classical gauge fields ($\tau \lesssim 0.2$ fm), (2) kinetic theory of HTL resummed gluons and quarks, (3) viscous hydrodynamics of the QGP, and (4) kinetic theory of hadrons.

When the constituents of matter are weakly coupled, such as is the case in gases, the matter transport can be described by time-dependent occupation probabilities $f(x,p,t)$ of (quasi-)particle modes in phase space, neglecting rapidly evolving quantum mechanical phases. The standard evolution equation for dilute systems of this type is the Boltzmann equation, which describes the evolution in terms of free streaming of (quasi-)particles between isolated rare collisions. 

The Boltzmann equation provides an adequate description of transport phenomena when the mean free path of the constituents of the medium is large compared to the distance between particles. When the mean free path is short or when correlations among the medium constituents are important, such as in liquids or other strongly coupled materials, other methods need to be employed. If the density and flow gradients in the liquid are small relative to the size of the system and the system is near thermal equilibrium, its motion can be described by viscous fluid dynamics, which locally tracks the evolution of energy, momentum and other locally conserved quantum numbers, such as electric charge and baryon number. A dimensionless criterion for the applicability of fluid dynamics is provided by the Knudsen number ${\rm Kn} = \lambda/L$, where $\lambda$ is the mean free path and $L$ denotes the system size. The applicability of fluid dynamics requires ${\rm Kn} \ll 1$, where the precise threshold depends on details of the interaction among the medium constituents.

Because the quark-gluon plasma (QGP) was originally conceived as the high-temperature phase of hadronic matter based on the property of asymptotic freedom, i.~e.\ the weakening of the QCD coupling constant $\alpha_s$ at short distances, the QGP was widely envisioned as a weakly coupled state of matter with the properties of a moderately dilute gas. Data from the first Au+Au run at RHIC immediately showed this to be a misconception as they revealed various striking patterns of collective flow \cite{Teaney:2000cw,Kolb:2000fha,Kolb:2003dz,STAR:2000ekf,STAR:2002hbo,PHENIX:2002hqx}. Since then a wealth of relevant data has been accumulated from multiple symmetric nuclear collision systems ($^{238}$U, $^{208}$Pb, $^{197}$Au, $^{129}$Xe, $^{96}$Ru, $^{96}$Zr, $^{63}$Cu) over a wide range of collider energies in the range $7.7~{\rm GeV} \leq \sqrt{s_{\rm NN}} \leq 5.36~{\rm TeV}$. The data cover inclusive and particle-identified spectra of hadrons and their azimuthal angular distributions with respect to the event-by-event collision plane, as well as multiparticle angular correlations.

The interpretation of these data is based on increasingly sophisticated simulations of the collision process in terms of second-order viscous relativistic fluid dynamics for the QGP phase augmented by hadronic Boltzmann transport following the hadronization of the QGP \cite{Teaney:2003kp,Romatschke:2007mq,Luzum:2008cw,Song:2010mg,Schenke:2010rr,Gale:2013da,Shen:2014vra}.  The input into these simulations are event-by-event fluctuating initial conditions for the fluid dynamics and various bulk properties of the QGP, including  the equation of state and the shear and bulk viscosities. Notwithstanding some conceptual problems of this approach during the earliest stages of the collision (see e.~g.~\cite{Plumberg:2021bme}, the results obtained in these simulations after appropriate choice of initial conditions and transport parameters, especially the kinematic shear viscosity $\eta/s$, reveal an impressive agreement with the data at collision energies of $\sqrt{s_{\rm NN}} = 200~{\rm GeV}$ and above.)

In recent years extensive Bayesian analyses of the optimal values for the parameters controlling the fluid dynamics simulations and their uncertainties have been performed
\cite{Bernhard:2016tnd,Bernhard:2019bmu,JETSCAPE:2020mzn,Nijs:2020roc}. The estimates for the temperature dependent kinematic shear viscosity lie in the range $0.1 \leq \eta/s \leq 0.2$, which is consistent both with the NLO prediction from HTL perturbation theory \cite{Ghiglieri:2018dib} for $g=2$ ($\eta/s \approx 0.136$) and the NLO holographic prediction from ${\cal N}=4$ SYM theory for the equivalent 't Hooft coupling $\lambda = g^2N_c = 12$ ($\eta/s = 0.114$). While the latter value cannot be directly compared with experiment -- the SYM theory differs in important aspects from QCD \cite{Huot:2006ys} -- the convergence of the values from weak and strong coupling approaches indicates that the in-medium gluon and quark modes in the QGP are strongly coupled and rather short-lived. The perturbative width of the thermal gluon inside the QGP at this coupling is $\gamma_{\rm g} = (0.9-1.1) T$ over the entire range of thermal momenta, which means\footnote{Note that $\gamma_g$ represents the imaginary part of the pole in the gluon propagator, which is half of the decay width of the mode intensity.} that the survival probability of a thermal gluon inside the QGP decays as $e^{-2\gamma_{\rm g}t} \sim e^{-2Tt}$. Whether calling such a short-lived mode with mean thermal energy $\langle\omega\rangle = 3T$ and width $\Gamma\approx 2T$ a ``quasiparticle'' stretches the boundaries of this notion beyond its utility is a matter for debate.

Figure~\ref{fig:Transport} shows NLO HTL predictions for two important transport coefficients of the QGP as a function of the gauge coupling $g$: The kinematic shear viscosity $\eta/s$ (left panel) and the heavy quark diffusion constant $2\pi TD_s$. (right panel) Both quantities are shown as solid lines; in comparison, both figures show results from the strongly coupled ${\cal N}=4$ SYM theory. The figures show that NLO perturbative predictions for these transport coefficients reach the strong coupling regime at realistic couplings $g \approx 2$ and are consistent with determinations  from Bayesian data analysis. (For a discussion of the diffusion constant $D_s$ see Section \ref{sec:Hard}.)

\begin{figure}[hbt]
\centering
	\includegraphics[width=0.45\linewidth]{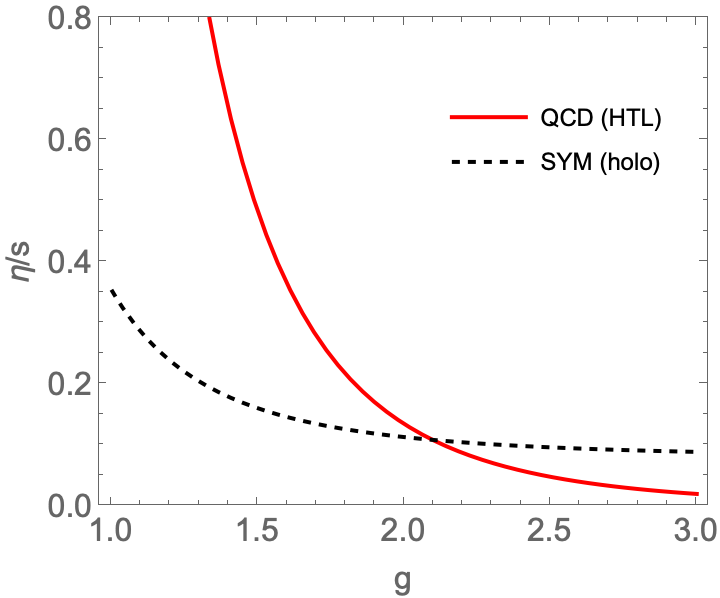}
	\hspace{0.03\linewidth}
	\includegraphics[width=0.45\linewidth]{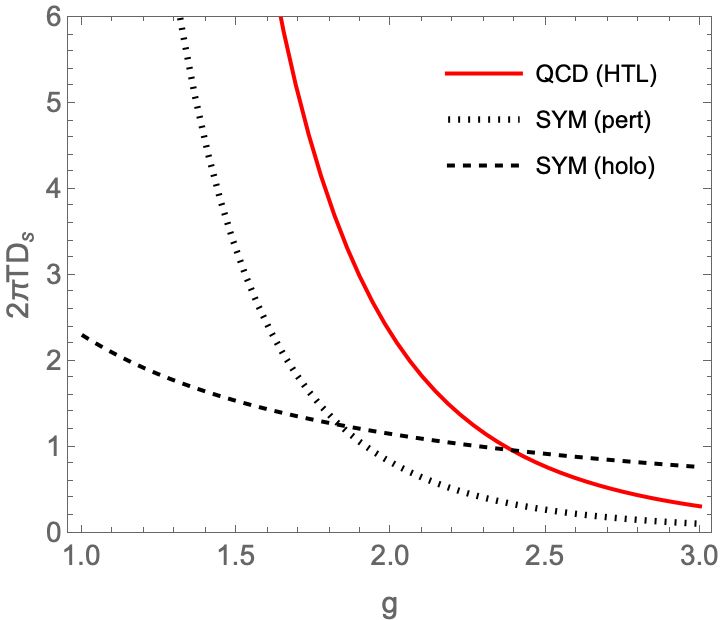}
\caption{{\it Left panel:} The NLO perturbative result for the kinematic shear viscosity $\eta/s$ of the QGP \cite{Ghiglieri:2018dib} is shown as a function of the gauge coupling $g$ by a solid red line. For comparison, the NLO result for the strongly coupled ${\cal N}=4$ SYM theory \cite{Buchel:2008sh} is shown as a dashed line (the weak coupling result at NLO is not known for the SYM theory.) The two lines intersect near $g=2$ ('t Hooft coupling $\lambda=12$) at a common value $\eta/s \approx 0.11$, which is compatible with values derived from experimental data by means of Bayesian analysis. {\it Right panel:} The normalized spatial diffusion constant $2\pi TD_s$ for slowly moving heavy quarks is shown as a function of the gauge coupling $g$ by a solid red line. $D_s$ is related to the drag coefficient $\gamma_D$ by $D_s=T/(M\gamma_D)$ and to the momentum fluctuation constant $\kappa$ by $D_s=2T^2/\kappa$. Results for the ${\cal N}=4$ SYM theory are shown in comparison: The NLO weak coupling result for SYM \cite{Caron-Huot:2007rwy} is shown as a dotted line, and leading-order strong coupling result \cite{Casalderrey-Solana:2006fio} is shown as a dashed line. The NLO perturbative results are seen to reach the strong coupling regime in the range $2 \lesssim g\lesssim 2.5$.}
\label{fig:Transport}
\end{figure}

One way to probe for the longevity of possible quasiparticles that constitute the bulk of the QGP is to study the system size dependence of the properties of the QGP. If the quasiparticles are short-lived, they cannot travel long distances and thereby explore the size of the QGP fireball.  At physical coupling ($g\approx 2, \alpha_s\approx 0.3$) the gluon half-width in the thermal momentum range is $\gamma\approx T$, which means that a gluon ``lives'' only for a very short time $\tau_{\rm g} \approx 1/(2T) \approx 0.3$ fm/c. This means that a QGP of even just $R = 1$ fm radius could have nearly the same transport properties as a much larger QGP fireball. Assuming a temperature $T \approx 300$ MeV and a boost invariant longitudinal expansion at proper time $\tau \approx R$ this corresponds to an entropy per unit rapidity of $dS/dy \approx \pi (RT)^3 s(T) \approx 160$ for $s(T)/T^3\approx 15$ \cite{Borsanyi:2013bia} or a final-state charge particle multiplicity $dN_{\rm ch}/dy \approx 25$. Collective flow in such small collision systems that is virtually identical to that in large system, such as Au+Au or Pb+Pb, has indeed been widely reported \cite{CMS:2010ifv,ALICE:2012eyl,CMS:2018loe} and even shown to be responsive to the collision geometry \cite{PHENIX:2018lia}. While these results do not provide evidence for the existence of quasiparticles in the QGP other than phonons, they do lend support to the notion that the constituent quasiparticles must be short-lived on a time scale of substantially less than 1 fm/c.

What else can measurements of bulk particle emission tell us about the internal structure of the QGP beyond the conclusion that its constituent quasiparticles are moderately strongly coupled and, by inference, have a lifetime of the order of the thermal length? One of the most remarkable features of the anisotropic collective flow pattern of different hadrons is that it obeys the constituent quark number scaling law \cite{PHENIX:2006dpn,STAR:2007afq,PHENIX:2007tef,STAR:2008ftz,STAR:2015gge,ALICE:2018yph} $v_k^{(\alpha)}(p_T) = n_{\rm q}^{(\alpha)}\tilde{v}(p_T/n_{\rm q,\alpha})$ where $k=2,3,\ldots$ is the order of the flow Fourier component, $\alpha$ denotes the hadron species, $\tilde{v}_k(p_T)$ is the same function for all light hadrons, and $n_{\rm q,\alpha}=2,3$ is the number of constituent (valence) quarks in the hadron. Near threshold, the scaling law works even better if the transverse momentum $p_T$ is replaced with the transverse kinetic energy $\sqrt{p_T^2+m^2}-m$, which better captures the kinematics of hydrodynamic flow at low momentum \cite{STAR:2008ftz}. The most natural explanation of the pervasive existence of this scaling law is that hadrons are formed by coalescence of quarks and antiquarks from the QGP \cite{Molnar:2003ff,Fries:2003kq,Greco:2003mm}. The function $\tilde{v}_k(p_T)$ then can be interpreted as the $k$-th Fourier component of the collective flow  of quarks (and probably gluons) inside the QGP at the moment of hadronization. This does not mean that quarks in the QGP are ``good'', i.~e.\ long-lived, quasiparticles, it only implies that quarks are deconfined in the QGP and individually participate in the collective flow \cite{Fries:2003vb}. 

Unfortunately, little direct evidence about gluons can be derived from bulk QGP transport in a similar way, because gluons do not have quantum numbers that survive hadronization. Indirect evidence of the presence of thermal gluons in the QGP comes from the production of strange quarks, because only the process $gg \to s\bar{s}$ is fast enough to  fully equilibrate the thermal phase space of strange quarks during the lifetime of the QGP \cite{Rafelski:1982pu,Koch:1986ud,Biro:1990vj}, as implied by the data from RHIC and LHC \cite{Becattini:2005xt,Becattini:2014hla}. In the HTL kinetic theory of quark flavor equilibration by gluons, chemical equilibrium of the QGP is attained on a time scale $\tau_{\rm chem} \sim 15(\eta/s)T^{-1}$ \cite{Kurkela:2018oqw}. For the value $\eta/s \approx 0.15$ derived from collective flow data, $\tau_{\rm chem} \approx 1.5$ fm/c, consistent with full chemical equilibration of all light quark flavors in the QGP.

\section{Hard Probes of the QGP}\label{sec:Hard}

Hard probes of the QGP include photons, heavy ($c,b$) quarks, and jets. Because they are either weakly interacting (photons) or characterized by a large scale (heavy quarks and jets), their production can be calculated perturbatively, and their modification in nuclear collisions due to interactions in the QGP can be calibrated with data from p+p collisions. We now discuss these probes beginning with jets.

\subsection{Jet Probes}\label{sec:Jets}

High-momentum jets in relativistic heavy-ion collisions usually hadronize outside the QGP. This implies that their propagation inside the QGP can be reliably described as branchings on the partonic level. The propagation and branching of jet partons are affected by the QGP in three main ways: 
\begin{itemize}
\item 
{\it Transverse momentum broadening:} The transverse momenta of jet partons are modified by interactions with the QGP. If the QGP acts as a system of quark and gluon quasiparticles, this interaction is described by the HTL broadening kernel \cite{Schlichting:2021idr,Moore:2021jwe} discussed in Section \ref{sec:HTL}. The average pathlength dependent momentum broadening due to interactions with the QGP is given by $\langle\Delta{\bf q}_\perp^2(L) = \hat{q}L$ where $\hat{q}$ is the jet quenching parameter \cite{Arnold:2008vd,Ghiglieri:2015ala}. In the strong coupling limit the interaction of the jet partons with the QGP fluid is dominated by the emission of phonons. Holographic calculation of the jet quenching parameter in the strongly coupled SYM theory yield values for $\hat{q}$ that are significantly larger than those obtained in HTL theory for the same 't Hooft coupling, even after corrections accounting for the different number of degrees of freedom are applied \cite{Liu:2006ug,Gubser:2008as,Casalderrey-Solana:2011dxg}.
\item
{\it Radiative energy loss:} Multiple scattering keeps jet partons longer off-shell and able to lose energy by emitting gluon radiation. The radiative energy loss in the QGP is suppressed by the Landau-Pomeranchuk-Migdal (LPM) effect resulting in a rate of energy loss that grows linearly with the pathlength in the plasma $dE/dL \propto -\alpha_s\hat{q}L$ \cite{Baier:1996kr}. This formula applies when the interaction with the QGP is dominated by multiple scattering. For situations where jet partons experience only a few scatterings, the radiative parton energy loss can be calculated in the opacity expansion \cite{Gyulassy:2000fs,Wiedemann:2000za}. In the strongly coupled limit of ${\cal N}=4$ SYM theory the rate of energy loss grows even more rapidly with the pathlength, $dE/dL \propto (E L^2)/x_{\rm stop}^3$ \cite{Chesler:2014jva,Chesler:2015nqz}, where $x_{\rm stop} \propto E^{-1/3}$ is the stopping distance for an energetic quark in the SYM plasma \cite{Gubser:2008as}. Because there are no propagating gluon modes in the strongly coupled SYM plasma, there is no distinction between radiative and collisional energy loss.
\item 
{\it Color decoherence:} As the jet partons interact with the QGP via gluon exchange, they exchange color with the medium. This effect destroys the color coherence among the jet fragments and removes the angular ordering of subsequent branchings that governs jet fragmentation in the vacuum \cite{Mehtar-Tani:2011hma}. The color screening length in the QGP plays an important role in this process as it governs how far a radiated gluon has to be transversely separated from its parent to act as an independent radiation source \cite{Casalderrey-Solana:2012evi}. This effect could, in principle, provide a mechanism for the determination of the color screening lnegth in the QGP.
\end{itemize}

An enormous wealth of data on QGP-modified jets has been collected from heavy-ion collisions especially at LHC and to a lesser extent at RHIC (see \cite{Busza:2018rrf,Harris:2023tti} for recent overviews and referencess). Observables of particular interest have been suppression factors $R_{\rm AA}(p_T)$ for single inclusive and idenitfied hadrons, for coincident back-to-back hadrons, inclusive jets and dijets, and modifications of jet fragmentation functions and jet shapes in heavy-ion collisions. In recent years, heavy-ion experiments at the LHC have increasingly focused on jet substructure observables. Energy-energy correlators have been proposed as especially sensitive and theoretically controlled measures of jet substructure \cite{Andres:2022ovj,Andres:2023xwr,Yang:2023dwc,Andres:2024ksi,Rai:2024ssx}, but data from relativistic heavy-ion collisions are just beginning to come forth \cite{CMS:2025ydi,Rai:2025qm}.

We are here not concerned with a general comparison of theoretical jet quenching calculations with the RHIC and LHC data, but with the question what these data tell us or can tell us about ``what the QGP is made of.'' Nuclear suppression factors for hadrons and jets inform us about the jet quenching parameter $\hat{q}$ or, for a low-opacity medium, about the path-integrated density of gluons and quarks in the QGP. The value deduced from multiple observables by means of a Bayesian analysis \cite{JETSCAPE:2024cqe} lies in the range $\hat{q}/T^3 = (8\pm 2)$ (see \cite{JET:2013cls} for early results using only inclusive hadron suppression data). This value is roughly consistent with the relation between $\hat{q}$ and $\eta/s$ that holds for a medium composed of screened quark-gluon quasiparticles \cite{Majumder:2007zh,Muller:2021wri}.

Since the internal evolution of the jet is a weak coupling process that is well described by perturbative branchings, any attempt to describe jet fragmentation within a strongly coupled gauge theory must fail. However, when the bulk QGP is strongly coupled, the energy deposition from the jet into the QGP could be amenable to a strong coupling approach. Such a hybrid model \cite{Casalderrey-Solana:2014bpa,Casalderrey-Solana:2016jvj} applies differential energy loss and transverse momentum broadening to a perturbatively generated jet shower. The energy loss and momentum broadening are modeled after strong coupling results \cite{Gubser:2008as,Arnold:2011qi,Chesler:2014jva,Chesler:2015nqz} but use data-adjusted values for the stopping distance $x_{\rm stop}$ of an energetic parton and for $\hat{q}$ \cite{Casalderrey-Solana:2018wrw}.\footnote{The fact that both, the $x_{\rm stop}^{-1}$ and $\hat{q}$ are too large when a realistic value of the 't Hooft parameter ($\lambda \approx 12$) is used, even after correction for the different number of degrees of freedom, suggests that this model may differ in an important detail of the plasma response to a perturbation that travels along the light-cone. It would be interesting to compare the structures of the broadening kernel $C(q)$ calculated in the HTL formalism \cite{Moore:2021jwe} with an analogous quantity in the ${\cal N}=4$ SYM theory at the same coupling.} The hybrid model can be used to simulate the energy and momentum deposition into the QGP that serves as the source for the hydrodynamic response of the QGP to the passage of the jet \cite{Casalderrey-Solana:2020rsj,Rai:2024ssx}. Models based on a quasiparticle description of the jet interaction with the QGP providing the source term for the hydrodynamic response give similar results \cite{Yang:2022nei}. (The hydrodynamic source term due to passage of a fast parton has been derived in HTL perturbation theory in \cite{Neufeld:2010xi}.) First evidence for the presence of such a jet wake in the backward direction has recently been found in $Z$-boson triggered jets in Pb+Pb collisions \cite{CMS:2024fli}. Systematic and detailed studies of such diffusion wakes could help explore the response of the QGP to localized injection of energy and momentum.

The proposal has been made to identify the scale $q_a$ of the transition from a quasiparticle structure of the QGP to a structureless liquid by looking for broadening effects on the internal substructure of the jet \cite{DEramo:2012uzl,DEramo:2018eoy}. Scatterings involving momenta $q \gg q_a$ would show up as a power law tail, whereas scatterings at $q \ll q_a$ would be governed by a Gaussian distribution. It is tempting to identify this scale with that introduced in Moli\`ere's theory of multiple scattering \cite{Moliere:1948zz,Bethe:1953va} as the scattering angle at which angular distribution due to scatterings on individual scattering centers, i.~e.\ quasiparticles in the QGP, transitions into a Gaussian distribution reflecting many random scatterings. The conceptual problem with this identification is that the transition between the two angular regimes in Moli\`ere theory depends only on the screening length of the single-center scattering kernel $C(q)$ and on the pathlength in the medium. It does not distinguish between a gaseous, liquid, or solid structure. Nevertheless, a measurement of the Moli\`ere screening angle would shed light on the color screening length in the QGP, which could inform models of the internal structure of the QGP. One recent measurement of the angular substructure of jets in heavy-ion collisions failed to find evidence for a power-law tail in the interactions between jet partons and the QGP that would indicate the presence of localized quasiparticles in the QGP \cite{ALICE:2024fip}. Other experimental approaches that quantify the changes in the angular substructure of photon-tagged jets have revealed the difficulty of such measurements due to implied selection biases \cite{CMS:2024zjn}.

\subsection{Heavy Quark Probes}\label{sec:Quarks}

Heavy quarks (charm and bottom) are able to probe the structure of the QGP via the diffusion constant and the (color) screening length. The ``melting'' of heavy quarkonium states due to color screening \cite{Matsui:1986dk,Karsch:1990wi} was one of the most intensely pursued signatures of QGP formation in the discovery period (1995--2005), and a rich phenomenology of sequential suppression of charmonium and Upsilon states has been established (see e.~g.\ \cite{Tang:2020hzi}). The theory of quarkonium ``melting'' in the QGP has also been greatly refined, and the importance of thermal ionization by gluons and possible recombination during QGP hadronization has been elucidated (see \cite{Rothkopf:2020vfz,Andronic:2024oxz} for reviews). Although this theoretical description has been constructed on the basis of the HTL quasiparticle picture of the QGP, it is not clear that this feature is essential as color screening also occurs in strongly coupled holographic models \cite{Bak:2007fk,Finazzo:2014zga}. In recent years, it has become increasingly clear that static color screening plays virtually no role in the disappearance of quarkonium states in the QGP, contrary to the original expectations \cite{Matsui:1986dk}. Lattice QCD simulations show that the real part of the quark-antiquark potential, reflecting static screening, is independent of the temperature, while the imaginary part, reflecting the thermal ionization rate, grows rapidly \cite{Bazavov:2023dci}. This result is qualitatively, but not quantitatively, consistent with the predictions from HTL perturbation theory \cite{Burnier:2007qm,Burnier:2013fca}.

Owing to their large mass, the momenta of heavy quarks usually far exceed the momentum exchanged with QGP constituents in a single interaction. This means that the effect of the QGP on a heavy quark can be treated as momentum diffusion, unless the energy of the heavy quark is very high, in which case radiative energy loss becomes important \cite{Djordjevic:2003zk,Djordjevic:2006tw}. In kinetic theory, the heavy quark diffusion constant $D_s$ measures the coupling between a slow-moving heavy quark and the medium.\footnote{Whereas the equation of state and the kinematic shear viscosity are solely properties of the medium itself, the jet quenching parameter $\hat{q}$ and the diffusion constant $D_s$ measure the strength of the coupling between the medium and an external probe (a high-energy parton or a heavy quark, respectively).} Its determination by comparing solutions of the Fokker-Planck equation \cite{Svetitsky:1987gq} or Langevin equation \cite{Caron-Huot:2007rwy} with experimental data thus provides sensitive information about the coupling strength of the QGP \cite{Scardina:2017ipo,He:2022ywp}. The data from relativistic heavy ion collisions for charm quarks are consistent with values in the range $3\lesssim 2\pi TD_s \lesssim 5$ and point to a strongly coupled QGP ($g \approx 2$).

\subsection{Electromagnetic Probes}\label{sec:Photons}

Photons and lepton pairs stand out among other probes of the QGP as being primarily sensitive to the presence of quarks and their quasiparticle structure. The photon emissivity of the QGP has been calculated in HTL perturbation theory up to next-to-leading order \cite{Ghiglieri:2013gia}. It turns out that several large NLO contributions almost cancel, resulting in $O$(10\%) corrections in the momentum range of interest $p_\perp \geq 5T \approx 1.5$ GeV/c even for realistic QCD coupling $\alpha_s \approx 0.3$ ($g \approx 2$). Because they depend mostly on the high-momentum structure of the QGP, thermal photons in the experimentally accessible range are only indirectly sensitive to soft radiative interactions within the QGP, as opposed to transport coefficients such as the shear viscosity $\eta$, the jet quenching parameter $\hat{q}$, and other diffusion constants, which are highly sensitive to infrared physics. 

Experimental data on thermal photon radiation in Pb+Pb collisions at LHC are compatible with theoretical predictions for a flowing QGP containing quarks as quasiparticles \cite{Bailhache:2024evo}. Some recent theoretical model studies conclude that the thermal photon spectrum is strongly influenced by blue-shift effects from collective flow of the QGP and less sensitive to the QGP temperature \cite{Shen:2013vja,Massen:2024pnj}. Other investigations argue that the slope of the thermal photon spectrum is sensitive to the temperature at which the QGP is formed \cite{Paquet:2022wgu}. Measurements of the photon yield in Au+Au collisions at RHIC \cite{PHENIX:2014nkk,STAR:2016use} agree on the centrality dependence, but not on the absolute magnitude. Thermal dilepton measurements from the QGP, while less sensitive to QGP flow than photon spectra, are rather difficult because dileptons from the QGP are overshadowed by hadronic backgrounds including charm decays. Dilepton spectra at invariant masses below the $\rho-\omega$ meson peak are mostly sensitive to the hadronic final state and reflect the thermal conditions at hadronization. Dilepton measurements in the intermediate invariant mass range 1 MeV $< M_{\rm ll} <$ 3 GeV have been reported both from the SPS \cite{NA60:2008ctj} and from RHIC \cite{STAR:2024bpc}, which are consistent with emission from a thermal source with temperature $T>T_c$. Recent calculations of dilepton production in nuclear collisions confirm that measurements in the intermediate mass range can be good probes for the formation temperature of the QGP \cite{Churchill:2023zkk} if the backgrounds can be sufficiently suppressed.

Overall, the most conservative conclusion at this time is that the electromagnetic probes of the QGP do not contradict the picture of a strongly coupled fluid composed of collectively flowing quarks and gluons, but the experimental data are not robust enough to permit conclusions about ``what the QGP is made of'' except that it contains collectively flowing quarks with approximately the expected thermal equilibrium density.

\section{Summary and Outlook}\label {sec:Outlook}

We have come a long way from the days of the January 2000 {\it Assessment of the Results from the CERN Lead Beam Programme} which described the quark-gluon plasma as a state in which ``quarks and gluons would freely roam within the volume of the fireball created by the collision'' \cite{Heinz:2000bk}. In fact, the developments in thermal perturbation theory have provided strong arguments for the conclusion ``that the notion of almost free gluons (and quarks) in the high-temperature phase of QCD is quite far from the truth'' \cite{Muller:1994rb}. But it was only after the compelling evidence for near-inviscid collective flow and strong jet quenching revealed by the experiments at RHIC that the notion that the QGP is a strongly coupled plasma became widely accepted \cite{Gyulassy:2004zy}.  At the same time, the qualitative success of strong coupling models based on the AdS/CFT conjecture and the abject failure of predictions from lowest-order HTL perturbation theory raised serious doubts that the bulk properties of the QGP could be understood on the basis of quark and gluon quasiparticles (see e.~g.\ \cite{Shuryak:2008eq}). 

Over the past decade, next-to-leading order calculations of dynamical processes in, and thermodynamic processes of, the fully developed QGP, away from the chiral/deconfinement crossover transition, have made it plausible that the QGP is, after all, a strongly coupled plasma composed of massive, very short-lived quarks and gluon quasiparticles. The essential new feature of NLO HTL perturbation theory is that collision rates at low momentum transfer are dominated by radiative processes which increase the final-state phase space and vastly enhance the efficiency of momentum redistribution in quasiparticle scattering. As already mentioned in Section \ref{sec:HTL}, these radiative processes give rise to an effective kinetic theory (``AMY'') that treats collinear ($1 \leftrightarrow 2$) splittings on an equal footing with binary ($2 \leftrightarrow 2$) scatterings \cite{Arnold:2002zm}. The AMY formalism extends the earlier concept of a parton cascade with collinear branchings at the leading logarithmic level \cite{Geiger:1991nj} into a full set of kinetic equations that include in-medium modifications of the quark and gluon quasiparticles together with detailed balance of forward and backward reactions. The AMY formalism has been used successfully to describe the rapid hydrodynamization and equilibration of the gluonic ``glasma'' \cite{Lappi:2006fp,Gelis:2012ri} that is thought to form the earliest form of matter at midrapidity in relativistic heavy ion collisions \cite{Kurkela:2018vqr,Kurkela:2018wud,Kurkela:2018oqw}.

As discussed in Section \ref{sec:HTL}, the HTL perturbative QGP has the remarkable property of sustaining a well developed phonon mode describing sound waves in the QGP at physical coupling $\alpha_s \approx 0.3$. This phonon mode emerges due to the very low kinematic shear viscosity $\eta/s$ of the QGP predicted by HTL perturbation theory when radiative processes are included at NLO. The rapid approach to this low-viscosity hydrodynamic description naturally arises from the AMY effective kinetic theory \cite{Heller:2016rtz}, thus completing the picture of the QGP as a liquid plasma composed of very short-lived, massive quark and gluon quasiparticles that rapidly emerges from an initial state characterized by dense gluon fields \cite{Lappi:2006fp,Gelis:2012ri,Schenke:2012wb}.

Remarkably, a similar picture arises in the ${\cal N}=4$ SYM theory model of a strongly coupled gauge theory with a gravity dual at NLO in the strong coupling expansion at 't Hooft coupling $\lambda = g^2N_c$ in the regime corresponding to the physical parameters ($g \approx 2, N_c=3$) of QCD \cite{Casalderrey-Solana:2018rle}. This convergence suggests that the physical properties of the QGP are general features of a gauge plasma at an intermediate coupling strength in the transition between the truly weak and strong coupling. It would be interesting, albeit technically difficult, to explore whether this apparent convergence persists at higher orders of perturbation theory in the ${\cal N}=4$ SYM theory where rigorous calculations are possible in both weak and strong coupling perturbation theory, similar to what was accomplished at NNLO for the equation of state \cite{Andersen:2021bgw},

There are several other interesting questions where further clarity would be desirable. One concerns the plasma instabilities that are known to exist in a longitudinally expanding QGP at weak coupling \cite{Mrowczynski:1993qm,Rebhan:2004ur,Arnold:2005ef} as well as in the expanding glasma \cite{Romatschke:2006nk}. Plasma turbulence driven by such instabilities can govern the dynamics of QGP thermalization \cite{Berges:2013eia}. In the hydrodynamic regime, the plasma turbulence driven by the longitudinal expansion can modify the shear viscosity of the QGP \cite{Asakawa:2006tc} and other QGP transport properties \cite{Romatschke:2004au}.

Another intriguing question is what defines a ``liquid'' in the relativistic domain where particle number is not conserved and interactions at short distances are not characterized by strong repulsion. Particle number conservation and short-distance repulsion form the foundation of the standard theory of nonrelativistic liquids \cite{Weeks:1971role,Hansen:2013theory}. Is there a viable theory of liquids that does not require a strongly repulsive interaction at short distances? One possibility is that absorptive interactions could play a similar role by reducing the probability of multiple particles to occupy the same volume. Such a mechanism is realized in QCD as two gluons occupying the same position in phase space can fuse into one, which leads to saturation of the gluon distribution in nuclei at small Bjorken-$x$ \cite{Gribov:1983ivg,Mueller:1985wy,Jalilian-Marian:1996mkd}. It is well known from the optical model of the nucleus that a (strongly absorptive) potential with a large imaginary part can suppress the scattering wave function at short distances (see e.~g.\ \cite{Austern:1961opt}).

Future measurements that could further buttress (or modify) our current notion of ``what the QGP is made of'' include:
\begin{itemize}
\item 
Precision measurements of the momentum spectra and azimuthal anisotropies of open heavy quark hadrons, especially $D$ and $B$ mesons, and the broadening of the azimuthal angular correlation between $D$-$\bar{D}$ pairs in heavy ion collisions \cite{Nahrgang:2013saa,Cao:2015cba} away from the back-to-back configuration predicted by perturbative QCD. These measurements would allow to better pin down the heavy quark spatial diffusion coefficient $D_s$ and the thermalization time for heavy quarks $\tau_Q=(M_Q/T)D_s$ \cite{ALICE:2022wwr}. 
\item
Extending the QGP enhancement effects for multistrange baryons, the deconfinement and thermalization of charm quarks in the QGP can be probed at LHC energies with the measurement of the yields of baryons containing multiple charm valence quarks ($\Xi_{cc},\Omega_{cc},\Omega_{ccc}$ \cite{ALICE:2022wwr}), which are predicted to be enhanced in Pb+Pb collisions by a factor up to $10^3$ compared with linear extrapolations from proton-proton collisions due to quark recombination \cite{Andronic:2021erx} if charm quarks thermalize. 
\item
More detailed measurements of the systematics of collective flow as a function of system size. Such measurements can involve medium size nuclei, such as $^{16}$O \cite{ALICE:2021wim,Brewer:2021kiv}, and nuclei with unusual geometric shape, such as $^{20}$Ne \cite{Giacalone:2024ixe}.
\item
While the valence quark scaling for the elliptic flow provides compelling, albeit indirect evidence that quarks in the QGP are unconfined, they do not prove that quarks in the QGP are quasiparticles, nor do they show that their chiral symmetry breaking constituent mass is absent. The most compelling evidence for the restoration of spontaneously broken chiral symmetry in the QGP could come from dileptons, which probe the photon spectral function \cite{Hohler:2015mda}. Chiral symmetry restoration manifests itself in a degeneracy of the vector and axial vector correlation functions, often referred to as $\rho$-$a_1$ mixing, and the convergence of the $\rho$ and $a_1$ masses at temperatures $T>T_c$ (see e.~g.\ Fig.~13 in \cite{Jung:2019nnr}). A precise measurement of the dilepton spectrum in the range 1 GeV $<m_{ee}<$ 2 GeV is a central goal of the ALICE-3 upgrade \cite{ALICE:2022wwr}. 
\item
Precision determinations of the parameters $\hat{q}$ and $\hat{\varepsilon}$ that control the radiative and collisional energy loss of energetic partons in a QGP \cite{Qin:2012fua} are sensitive to the in-medium masses of quark and gluon quasiparticles \cite{Coleman-Smith:2012bcs}. Disentangling these two mechanisms of energy loss, if possible, requires precise differential studies of leading hadron suppression for jets initiated by light and heavy quarks \cite{Djordjevic:2003zk,Djordjevic:2006tw} as they will be performed during LHC runs 3 and 4, as well as by sPHENIX at RHIC. However, the distinction between radiative and collisional energy loss becomes more and more ambiguous with increasing coupling and utterly meaningless in the strong coupling limit.
\item
High statistics studies of the angular decorrelation of dijets or of the internal angular structure of jets could, in principle, provide evidence for the presence of localized scattering centers in the QGP and, eventually, about their properties \cite{DEramo:2018eoy}. Measurements of the changes in the angular distribution of subjects in heavy ion collisions could yield information about the color resolution length in the QGP \cite{ATLAS:2025lfb}. QGP modifications to energy-energy correlators (EEC), which allow for better theoretical control than other jet observables, may provide alternative measurements of properties of the QGP, such  the jet quenching parameter $\hat{q}$ and the color screening length in the QGP  \cite{Andres:2022ovj,Andres:2023xwr,Andres:2024ksi}.
\item
Because of its very low kinematic viscosity, the energy and momentum injected into the QGP by a passing jet is predicted to lead to a collective response. Such a diffusion wake is a universal phenomenon at strong and moderate coupling when \cite{Gubser:2007ni,Neufeld:2008hs,Betz:2008ka}. First observations of this wake \cite{CMS:2024fli} are consistent with theoretical expectations providing confirmation of the notion that the QGP is an excellent fluid. Further studies of the wake have the potential to reveal useful information about the hydrodynamic response of the equilibrated QGP to a localized injection of energy and momentum \cite{Cao:2020wlm,Yang:2022nei,Kudinoor:2025ilx,Barata:2025fzd}.
\end{itemize}
These experimental opportunities and the associated theoretical challenges instill confidence in the expectation that our understanding of ``what the QGP is made off'' will continue to progress rapidly over the next decade and beyond.
\bigskip

{\it Acknowledgments:} I thank J.-F. Paquet for helpful discussions about photon emission from the QGP, J. Pawlowski for help in assessing the insights from the functional renormalization group approach, and S.A. Bass, K. Boguslavski, C. Gale, J.W. Harris, B.V. Jacak, J. Schenke, M. van Leeuwen, and W.A. Zajc for valuable comments on a draft of this manuscript. I am especially indebted to K. Rajagopal for his insightful comments on many aspects covered in this review. This work was supported by a grant from the U.~S.~Department of Energy, Office of Science (DE-FG02-05ER41367).

\bibliography{QGPat50}

\end{document}